\documentclass[aps,pra,twocolumn,groupedaddress,showpacs]{revtex4}

\usepackage{graphicx}
\usepackage{dcolumn}
\usepackage{bm}

\begin{document}


\title{Discovery of dumbbell-shaped Cs$^{\ast }$He$_{n}$ exciplexes in
solid $^{4}$He}


\author{D. Nettels}
\author{A. Hofer}
\author{P. Moroshkin}
\email[]{peter.moroshkin@unifr.ch}
\author{R. \surname{M\"{u}ller-Siebert}}
\author{S. Ulzega}
\author{A. Weis}

\affiliation{D\'{e}partement de Physique, Universit\'{e} de Fribourg,
Chemin du Mus\'{e}e 3, 1700 Fribourg, Switzerland}
\homepage[]{www.unifr.ch/physics/frap/}

\date{\today}

\begin{abstract}
We have observed several new spectral features in the fluorescence
of cesium atoms implanted in the hcp phase of solid helium
following laser excitation to the 6$^{2}$P states. Based on
calculations of the emission spectra using semiempirical Cs-He
pair potentials the newly discovered lines can be assigned to the
decay of specific Cs$^{\ast }$He$_{n}$ exciplexes: an apple-shaped
Cs$(A\Pi _{3/2})$He$_{2}$ and a dumbbell-shaped Cs$(A\Pi _{1/2})
$He$_{n}$ exciplex with a well defined number $n$ of bound helium
atoms. While the former has been observed in other enviroments, it
was commonly believed that exciplexes with $n>2$ might not exist.
The calculations suggest Cs$(A\Pi _{1/2}) $He$_{6}$ to be the most
probable candidate for that exciplex, in which the helium atoms
are arranged on a ring around the waist of the dumbbell shaped
electronic density distribution of the cesium atom.
\end{abstract}


\maketitle


Alkali atoms and helium atoms in their ground states strongly
repel each other by virtue of the Pauli principle. However, an
alkali atom excited to one of its $P$ states can exert an
attractive potential on a helium atom that can lead to bound
states, known as exciplexes. The formation of alkali-helium
exciplexes was considered for the first time by Dupont-Roc \cite
{DupontRoc;EXCITEDPSTATESLIQUIDHELIUM} and Karnorsky et al. \cite
{Kanorsky;PressureshiftHelium} as an explanation for the observed
quenching of atomic fluorescence from light alkali atoms (Na, Li)
embedded in liquid or solid $^{4}$He. In the meantime such
molecules have been observed in different environments, such as
liquid helium and cold helium gas \cite
{Yabuzaki;EmissionSpectraOfCsHeExcimer,Yabuzaki;EmissionSpectraOfRbHenExciplexes}
, as well as on the surface of helium nanodroplets \cite
{Stienkemeier;alkaliattachedtoheliumclusters,Reho;AlkaliExciplexForationI,Bruhl;RbHeexiFormationDroplets,Schulz;FormationExciplexesKHe}
. Here we present the first observations of such exciplexes in a
solid helium matrix.

In earlier experiments \cite {Kanorsky;PressureshiftHelium} we
have studied the excitation and fluorescence spectra of atomic
cesium implanted into the bcc and hcp phases of solid helium. It
was found that the excitation at the $D_{1}$ transition
(6S$_{1/2}$-6P$_{1/2}$) results in atomic fluorescence at the same
transition, blue shifted (with respect to the free Cs atom) by the
interaction with the helium matrix. At the same time, excitation
on the D$_{2}$ transition (6S$_{1/2}$-6P$_{3/2}$) produced merely
a weak fluorescence on the D$_{1}$ emission line, which indicates
that the 6P$_{3/2}$ atoms are partly quenched into the 6P$_{1/2}$
state, while the main relaxation channel remained unknown.
Recently, the extension of the spectral range of our detection
system allowed us to discover several new emission lines, red
shifted with respect to the atomic fluorescence line. We attribute
those lines to the formation and decay of Cs$^{\ast }$He$_{n}$
molecules.

The maximal number of helium atoms for different alkali-helium
exciplexes was found previously to be $n_{max}=4$ for K$^*$He$_n$
\cite {Schulz;FormationExciplexesKHe}, $n_{max}=6$ for
Rb$^*$He$_n$ \cite {Yabuzaki;EmissionSpectraOfRbHenExciplexes},
$n_{max}=2$ for Cs$^*$He$_n$ \cite
{Yabuzaki;EmissionSpectraOfCsHeExcimer}. Hirano et al. \cite
{Yabuzaki;EmissionSpectraOfCsHeExcimer} discuss on the basis of
Cs$^{\ast }$He$_{2}$-He potential energy surfaces that there
should be no stable Cs$^{\ast }$He$_{3}$ configuration. They
therefore conclude that exciplexes with more than $n_{max}=2$ do
not exist, since they regard the exciplex formation as a
sequential process (Cs$^{\ast }\rightarrow $Cs$^{\ast
}$He$_{1}\rightarrow ...\rightarrow $Cs$^{\ast }$He$_{n_{max}}$).
However, our experimental results demonstrate unambiguously that
in the hcp phase of solid helium Cs(A$\Pi _{1/2}$)He$_{n}$
exciplexes with $n>2$ are formed, when the cesium atoms are
excited to the 6P$_{3/2}$ state. From the relative integrated
observed line intensities we conclude that the formation of those
exciplexes is the most probable deexcitation channel of the
6P$_{3/2}$ state.

In the present experiment a $^{4}$He matrix doped with Cs atoms
was produced by the technique described in our earlier papers
\cite {Kanorsky;OptSpecAtomsTrappedSolidHe,Nettels;MultiPhoton1}.
Data were taken in the hcp phase of solid $^{4}$He at a
temperature of $1.5$~K and a pressure of $31.6$~bar. For the
excitation of the embedded atoms we used a single mode cw
Ti:Al$_2$O$_{3}$-laser, pumped by a Nd:YVO$_{3}$ laser. The laser
wavelength was tuned to the D$_{2}$ absorption line of cesium,
whose resonance wavelength is shifted to 800 nm due to the
influence of the surrounding helium matrix. The atomic
fluorescence light from the sample volume ($\sim 3$~mm$^{3}$) is
detected by a fiber coupled optical spectrum analyzer (Ando Co.
Ltd., AQ-6315A) which has a detection range of 350-1750~nm. The
spectral resolution was limited to about 5~nm.

\begin{figure}[tbp]
\includegraphics[width=8.0cm]{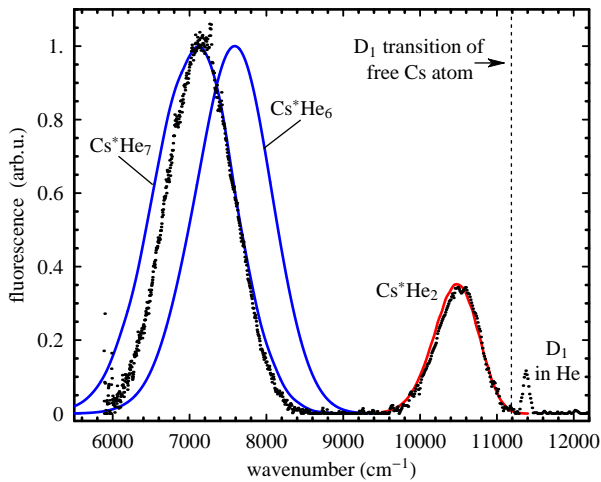}
\caption{Measured emission spectrum (dots) of matrix-isolated Cs
atoms in the hcp phase of solid $^4$He. The fluorescence emission
was observed following resonant D$_2$ excitation of the atoms. For
comparison also calculated emission lines (solid curves) of
Cs(A$\Pi _{3/2}$)He$_{2}$, Cs(A$\Pi _{1/2}$)He$_{6}$, and Cs(A$\Pi
_{1/2}$)He$_{7}$ exciplexes are presented.}
\label{fig:ExciplexSpectrum2}
\end{figure}

Fig.~\ref{fig:ExciplexSpectrum2} shows a typical recorded emission
spectrum, as well as calculated spectra of several Cs$^{\ast
}$He$_{n}$ exciplexes. The smallest peak at 11400~cm$^{-1}$
corresponds to fluorescence on the D$_{1}$ transition, indicating
the existence of a transfer channel between 6P$_{3/2}$ and
6P$_{1/2}$ states. That process was studied before, in experiments
with superfluid helium \cite
{OpitcalPropAlkaliAtomsPressSuperfluidHe;Kinoshita}. Another
emission line appears at 10520~cm$^{-1}$. We attribute this broad
and asymmetric peak to the emission of Cs($A\Pi _{3/2}$)He$_{2}$
exciplexes, which had previously been observed in liquid helium
and in cold helium gas \cite
{Yabuzaki;EmissionSpectraOfCsHeExcimer}. The peak presented here
is blue-shifted with respect to the one reported in  \cite
{Yabuzaki;EmissionSpectraOfCsHeExcimer} by approximately
500~cm$^{-1}$. The shift is due to the influence of the helium
matrix and increases with helium pressure. The detailed study of
that dependence in the bcc and hcp phases of solid helium will be
presented elsewhere \cite{Nettels;ExciplexesPhd}. There is yet
another prominent emission line at 7130~cm$^{-1}$, with a
linewidth of 1010~cm$^{-1}$ (FWHM), which is almost three times
stronger than the former peak. Our theoretical model suggests that
this newly discovered intense emission line can be assigned to a
higher order ($n>2$) exciplex.

\begin{figure}[tbp]
\includegraphics[width=7.8cm]{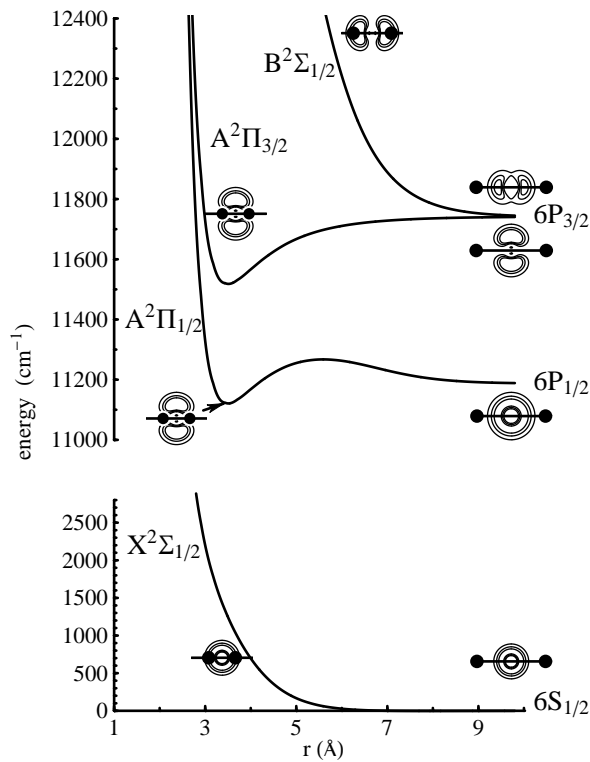}
\caption{Adiabatic potentials of the Cs$^{\ast }$He$_{2}$ system,
including the spin-orbit interaction. The two helium atoms are
located at $r$ and $-r$ on the quantization (rotationally
symmetry) axis, indicated in the pictographs by a solid line. The
shape of the electronic density distribution of the cesium atom
changes significantly as helium atoms (filled circles) approach.}
\label{fig:CsHe2PotentialPlot}
\end{figure}

\begin{figure}[t]
\includegraphics[width=7.8cm]{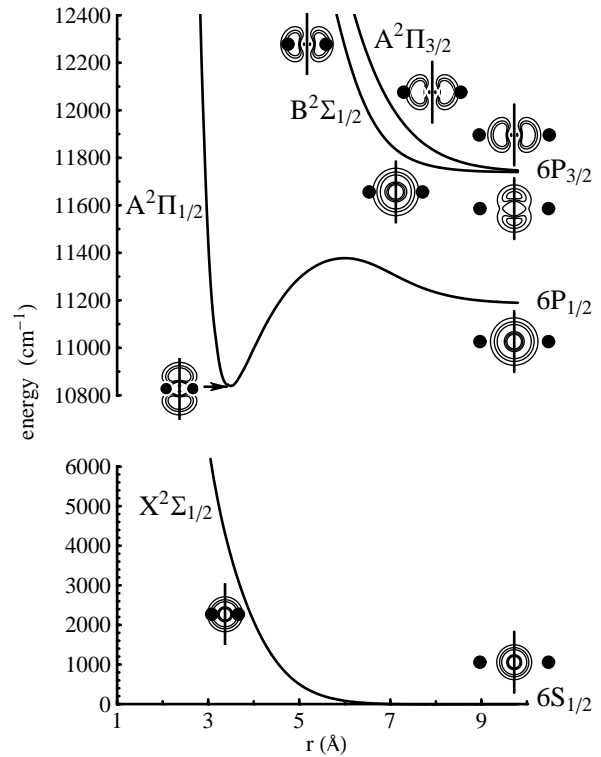}
\caption{Adiabatic potentials of the Cs$^{\ast }$He$_{6}$ system.
The six helium atoms are located on a ring of radius $r$
concentric with the symmetry axis. Only the $A^{2}\Pi _{1/2}$
potential has a binding attractive well.}
\label{fig:CsHe6PotentialPlot}
\end{figure}

We have calculated the emission spectra of Cs$^{\ast }$He$_{n}$
exciplexes with a treatment similar to those of \cite
{Yabuzaki;EmissionSpectraOfCsHeExcimer,Yabuzaki;EmissionSpectraOfRbHenExciplexes}.
We consider only the influence of the $n$ helium atoms that form
the exciplex and neglect the influence of the bulk of the
surrounding helium matrix. The interaction between one cesium atom
and the $n$ He atoms is described as the sum over adiabatic
molecular two-body interaction potentials. We use the pair
potentials between a helium ground state atom and alkali atoms in
their ground and lower excited states calculated by Pascale \cite
{Pascale;LDependentPseudoPotentialAlkaliHe}. For the 6S ground
state that potential, $V_{\sigma }^{6s}(r)$, is radially
symmetric. For the 6P states the interaction is anisotropic and
can be expressed by the operator
\begin{equation}
V^{6P}(\mathbf{r})=V_{\sigma }^{6P}(r)+\left(
\frac{\mathbf{L}\cdot \mathbf{r}}{\hbar r}\right) ^{2}[V_{\pi
}^{6P}(r)-V_{\sigma }^{6P}(r)], \label{eq:V6PCsHe}
\end{equation}
where $\mathbf{r}=\mathbf{r}(r,\theta ,\varphi )$ denotes the
position vector of a helium atom with respect to the cesium atom
and $\mathbf{L}$ is the orbital angular momentum operator of the
cesium valence electron \cite
{DupontRoc;EXCITEDPSTATESLIQUIDHELIUM}. Stable exciplexes of the
form Cs$^{\ast }$He$_{n=2}$ are formed by two helium atoms located
on a common axis on opposite sides of the cesium atom. For
Cs$^{\ast }$He$_{n\geq 3}$ the helium atoms are distributed on a
concentric ring around the alkali atom. The summation over the
pair potentials can be expressed by the operator
$V_{n}^{Cs\text{-}He}(r)=\sum_{i=1}^{n}V^{6P}(\mathbf{r}_{i}),$
with $\mathbf{r}_{i}=\mathbf{r}(r,\theta =\pi /2,\varphi
_{i}=i2\pi /n)$ and $n=1,2,...$. In addition we include He-He
interactions by summing over the corresponding
$V_{He\text{-}He}(R)$ potentials between neighboring helium atoms
using the semi-empirical potential given by Beck
\cite{Beck;InteratomicPotHeHe}. The distance $R$ between two
neighboring helium atoms is a function of the cesium--helium
separation $r$ and the number $n$ of helium atoms:
$R=|\mathbf{r}_{i}-\mathbf{r}_{i+1}|=2r\sin (\pi /n)$. After
including the spin-orbit interaction in Cs the total interaction
potential of the Cs$^{\ast }$He$_{n}$ system reads
\begin{equation}
V_{Cs^{\ast
}He_{n}}(r)=V_{n}^{Cs\text{-}He}(r)+nV_{He\,\text{-}\!He}(R)+2/3\Delta
\mathbf{L}\cdot \mathbf{S},  \label{eq:VCsHen}
\end{equation}
where $\Delta =554.0$~cm$^{-1}$ is the fine-structure splitting of
the free cesium 6P state and $\mathbf{S}$ the electronic spin
operator. $V_{Cs^{\ast }He_{n}}(r)$ is diagonalized algebraically.
In Figs.~\ref {fig:CsHe2PotentialPlot} and
\ref{fig:CsHe6PotentialPlot} the resulting $r$-dependencies of the
eigenvalues are shown for Cs$^{\ast }$He$_{2}$ and Cs$^{\ast
}$He$_{6}$ respectively. The same plots also show the ground state
potentials given by $nV_{\sigma
}^{6s}(r)+nV_{He\,\text{-}\!He}(R)$. The potentials are labelled
according to their electronic configurations as $X^{2}\Sigma
_{1/2}$, $A^{2}\Pi _{1/2}$, $A^{2}\Pi _{3/2}$ and $B^{2}\Sigma
_{1/2}$. The quantization axis is defined by the symmetry axis of
the exciplexes, which is the internuclear axis of the cesium atom
and the two helium atoms in the case of Cs$^{\ast }$He$_{n=1,2}$,
whereas for Cs$^{\ast } $He$_{n\geq 3}$ it is the axis of the
helium ring. Pictographs next to the curves show the variation of
the cesium electronic density as the $n$ helium atoms, indicated
by two filled circles, approach the cesium atom.

From the adiabatic potentials of
Figs.~\ref{fig:CsHe2PotentialPlot} and
\ref{fig:CsHe6PotentialPlot} one sees that the helium atoms, due
to the Pauli principle, are repelled by the cesium valence
electron. However, in cases where the atoms approach along a nodal
line or in a nodal plane of the electron distribution they
experience an attractive van der Waals force until they are
repelled by the cesium core. This definitely holds for the
Cs($A\Pi _{3/2}$)He$_{2}$ exciplex. For Cs($A\Pi _{1/2}$)He$_{2}$
the situation is more complicated. When the two helium atoms are
far away the electronic configuration is the one of the 6P$_{1/2}$
state of the free cesium atom, which has a spherical symmetry and
is hence, as the 6S$_{1/2}$ ground state, repulsive for the helium
atoms. However, when the helium atoms approach the cesium atom the
electronic wave function of the latter is deformed and becomes
apple-shaped. This new configuration now offers a binding
potential minimum. The formation of this second configuration has
a potential barrier of 79~cm$^{-1}$, which is much higher than the
thermal energy of $1.0$~cm$^{-1}$. If, on the other hand, the two
atoms do not approach simultaneously, but one after the other,
only the first atom has to overcome the potential barrier thereby
forming the apple-like electronic configuration of a Cs$^{\ast
}$He$_{1}$ structure, which then becomes attractive for a second
helium atom approaching from the opposite side. A third helium
atom approaching the Cs($A\Pi _{1/2}$)He$_{2}$ exciplex will again
be repelled. However, if it comes sufficiently close, the
electronic configuration changes to a dumbbell shape, which has a
binding minimum for the three helium atoms. Here too, a potential
barrier has to be overcome. The three atoms are then bound and
located on a ring concentric around the waist of the dumbbell. The
Cs($A\Pi _{1/2}$)He$_{3\text{ }}$structure exerts a purely
attractive potential on further helium atoms. With an increasing
number of helium atoms on the ring the repulsive potential between
those atoms increases, which puts a natural limit on the maximum
number $n_{\max }$ that can be accommodated. Only the $A\Pi
_{1/2}$ state can bind more than two helium atoms as can be seen
from Fig.~\ref{fig:CsHe6PotentialPlot}.

It was pointed out by Dupont-Roc
\cite{DupontRoc;EXCITEDPSTATESLIQUIDHELIUM} that the height of the
potential barrier is determined by the strength of the spin-orbit
interaction of the P state. If the L-S coupling is weak compared
to the alkali-helium interaction as in the case of Na, K, and Rb,
it can be neglected and the electronic configuration can be
approximated by a P$_z$ orbital that allows the formation of
dumbbell-shaped exciplexes with $n>2$. If on the other hand the
spin-orbit interaction dominates, as for cesium, one has to
consider the electron distributions of the L-S-coupled P$_{1/2}$
and P$_{3/2}$ states, which are spherical in the first case and of
apple shape in the second case, thereby restricting  $n_{max}$ to
2. In contrast to this simple model our treatment takes into
account that the electronic configuration changes adiabatically
its shape as helium atoms approach.

We calculated the adiabatic potentials and the vibrational
zero-point energies for all Cs$^{\ast }$He$_{n}$ systems up to
$n=9$. The treatment of the molecular vibrations is approximate
and will be described elsewhere. Higher vibrational states are not
populated at the temperature of the experiments and can be
neglected, as well as the contributions from rotations.

\begin{figure}
\includegraphics[width=7.4cm]{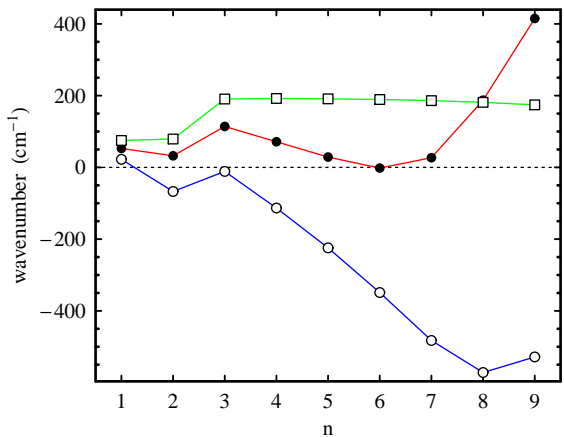}
\caption{Energy dependencies of the Cs$(A^{2}\Pi _{1/2})He_{n}$
exciplexes as a function of the number $n$ of helium atoms. Shown
are the minimal energies (open circles) of the potential wells,
the barrier heights (open squares) and the total zero-point
energies (filled circles). All energies are given with respect to
the dissociation limit, i.e., the electronic energy of the
6P$_{1/2}$ state. Corresponding points are joined by lines to
guide the eye. The temperature of the helium matrix corresponds to
$kT=1.0$~cm$^{-1}$.} \label{fig:energylevelplot}
\end{figure}

The $n$ dependence of the zero-point energies for Cs($A\Pi
_{1/2}$)He$ _{n}$ exciplexes is represented in
Fig.~\ref{fig:energylevelplot} by filled circles. Also shown are
the heights of the potential barriers (open sqares) and the depths
of the potential minima (open circles). All energies are given
with respect to the dissociation limit, i.e., the electronic
energy of the 6P$_{1/2}$ state. The significant increase of the
potential barrier when going from $n=2$ to $n=3$ reflects the
discussion given above. The configuration with $n=8$ has the
deepest attractive potential, while its zero-point energy due to
the stronger localization of helium atoms exceeds the barrier
energy, which makes that complex unstable.
Cs($A\Pi _{1/2}$)He$ _{7}$ seems to be the largest possible
quasi-bound exciplex. Only the molecule with $n=6$ has an energy
below the dissociation limit and is hence predicted to be stable.
Note that for rubidium several configurations are predicted to be
stable and Rb($A\Pi _{1/2}$)He$_{n}$ exciplexes with $n=1\ldots 6$
have been observed recently
\cite{Yabuzaki;EmissionSpectraOfRbHenExciplexes}.

In a last step we calculated the shape of the emission lines by
using the Franck-Condon approximation in a similar way as the
authors of \cite{Yabuzaki;EmissionSpectraOfCsHeExcimer}. Among all
exciplexes considered, the best agreement of the calculated
spectra with the experimental ones is provided by those shown in
Fig.~\ref{fig:ExciplexSpectrum2} by solid curves. The shape and
width of the line at 10520~cm$^{-1}$ agrees well with the
calculated emission line of Cs(A$\Pi _{3/2}$)He$_{2}$. However,
the apparently excellent agreement between the positions of these
lines is accidental as the pressure shift, i.e., the influence of
the surrounding helium matrix was not taken into account. In fact,
K. Enomoto et al. \cite{Yabuzaki;EmissionSpectraOfCsHeExcimer}
measured the emission line of Cs(A$\Pi _{3/2}$)He$_{2}$ exciplexes
in cold helium vapor environment to lie at about 10000~cm$^{-1}$.
We like to mention that that we have also observed fluorescence
from Cs(A$\Pi _{1/2}$)He$_{2}$ exciplexes, not discussed here,
following D$_1$ excitation
\cite{Nettels;ExciplexesPhd,Nettels;Phd}. Such exciplexes were
also observed recently on cesium doped He clusters
\cite{Stienkemeier;CsPi12He2}.

The correct assignment of the strongest measured peak at
7130~cm$^{-1}$ in Fig.~\ref {fig:ExciplexSpectrum2} is a more
subtle task. It is clear that the line originates from a Cs(A$\Pi
_{1/2}$)He$_{n>2}$ exciplex, in which the helium atoms form a ring
in the nodal plane of the Cs atom. It is very likely that this
line also shifts with helium pressure. Moreover, as the
ring-shaped exciplexes contain a larger number of helium atoms any
imprecision of the initial pair potentials will be amplified. As a
consequence the line positions of the higher order Cs$^{\ast
}$He$_{n}$ systems cannot be predicted with sufficient accuracy to
allow an unambiguous assignment of the observed line. The overall
good agreement between calculated and measured lineshapes
indicates that the line at 7130 cm$^{-1}$ originates from the
decay of an exciplex with a specific number $n$ of bound helium
atoms and that it is not a superposition of lines from exciplexes
with different values of $n$. Such superpositions were observed in
the case of Rb$^{\ast }$He$_{n}$ exciplexes in cold $^{4}$He vapor
\cite{Yabuzaki;EmissionSpectraOfRbHenExciplexes}. The theoretical
considerations suggest, as a best guess, that the observed peak at
7130~cm$^{-1}$ originates from decaying Cs$^{\ast }$He$_{6}$
complexes, as those exciplexes have minimal energy and are the
only ones predicted to be stable.

In summary, we have performed a study of laser induced
fluorescence of cesium atoms trapped in the hcp phase of a helium
crystal. We have reported the observation of new spectral
features, which are broader and more intense than the pure atomic
lines. We believe that those lines are formed by the emission from
two types of specific Cs$^{\ast }$He$_{n}$ exciplex structures,
viz., an apple-shaped complex with two helium atoms bound to the
Cs atom and a dumbbell-shaped complex, in which a ring of helium
atoms is bound to the nodal plane of the Cs wave function. These
assignments are supported by model calculations, which allow us to
obtain the corresponding emission spectra. In the case of the ring
structure the calculations suggest $n=6$ as the most likely number
of bound atoms.


We like to thank J. Pascale for sending us his numerical Cs-He
pair potentials. This work was supported by a grant of the
Schweizerischer Nationalfonds.

\bibliography{HePaper}

\begin{thebibliography}{16}
\expandafter\ifx\csname natexlab\endcsname\relax\def\natexlab#1{#1}\fi
\expandafter\ifx\csname bibnamefont\endcsname\relax
  \def\bibnamefont#1{#1}\fi
\expandafter\ifx\csname bibfnamefont\endcsname\relax
  \def\bibfnamefont#1{#1}\fi
\expandafter\ifx\csname citenamefont\endcsname\relax
  \def\citenamefont#1{#1}\fi
\expandafter\ifx\csname url\endcsname\relax
  \def\url#1{\texttt{#1}}\fi
\expandafter\ifx\csname urlprefix\endcsname\relax\def\urlprefix{URL }\fi
\providecommand{\bibinfo}[2]{#2}
\providecommand{\eprint}[2][]{\url{#2}}

\bibitem[{\citenamefont{Dupont-Roc}(1995)}]{DupontRoc;EXCITEDPSTATESLIQUIDHELI%
UM}
\bibinfo{author}{\bibfnamefont{J.}~\bibnamefont{Dupont-Roc}},
  \bibinfo{journal}{Z. Phys. B} \textbf{\bibinfo{volume}{98}},
  \bibinfo{pages}{383} (\bibinfo{year}{1995}).

\bibitem[{\citenamefont{Kanorsky et~al.}(1995)\citenamefont{Kanorsky, Weis,
  Arndt, Dziewior, and H{\"a}nsch}}]{Kanorsky;PressureshiftHelium}
\bibinfo{author}{\bibfnamefont{S.}~\bibnamefont{Kanorsky}},
  \bibinfo{author}{\bibfnamefont{A.}~\bibnamefont{Weis}},
  \bibinfo{author}{\bibfnamefont{M.}~\bibnamefont{Arndt}},
  \bibinfo{author}{\bibfnamefont{R.}~\bibnamefont{Dziewior}}, \bibnamefont{and}
  \bibinfo{author}{\bibfnamefont{T.~W.} \bibnamefont{H{\"a}nsch}},
  \bibinfo{journal}{Z. Phys. B} \textbf{\bibinfo{volume}{98}},
  \bibinfo{pages}{371} (\bibinfo{year}{1995}).

\bibitem[{\citenamefont{Enomoto et~al.}(2002)\citenamefont{Enomoto, Hirano,
  Kumakura, Takahashi, and Yabuzaki}}]{Yabuzaki;EmissionSpectraOfCsHeExcimer}
\bibinfo{author}{\bibfnamefont{K.}~\bibnamefont{Enomoto}},
  \bibinfo{author}{\bibfnamefont{K.}~\bibnamefont{Hirano}},
  \bibinfo{author}{\bibfnamefont{M.}~\bibnamefont{Kumakura}},
  \bibinfo{author}{\bibfnamefont{Y.}~\bibnamefont{Takahashi}},
  \bibnamefont{and} \bibinfo{author}{\bibfnamefont{T.}~\bibnamefont{Yabuzaki}},
  \bibinfo{journal}{Phys. Rev. A} \textbf{\bibinfo{volume}{66}},
  \bibinfo{pages}{042505} (\bibinfo{year}{2002}).

\bibitem[{\citenamefont{Hirano et~al.}(2003)\citenamefont{Hirano, Enomoto,
  Kumakura, Takahashi, and
  Yabuzaki}}]{Yabuzaki;EmissionSpectraOfRbHenExciplexes}
\bibinfo{author}{\bibfnamefont{K.}~\bibnamefont{Hirano}},
  \bibinfo{author}{\bibfnamefont{K.}~\bibnamefont{Enomoto}},
  \bibinfo{author}{\bibfnamefont{M.}~\bibnamefont{Kumakura}},
  \bibinfo{author}{\bibfnamefont{Y.}~\bibnamefont{Takahashi}},
  \bibnamefont{and} \bibinfo{author}{\bibfnamefont{T.}~\bibnamefont{Yabuzaki}},
  \bibinfo{journal}{Phys. Rev. A} \textbf{\bibinfo{volume}{68}},
  \bibinfo{pages}{012722} (\bibinfo{year}{2003}).

\bibitem[{\citenamefont{Stienkemeier et~al.}(1996)\citenamefont{Stienkemeier,
  Higgins, Callegari, Kanorsky, Ernst, and
  Scoles}}]{Stienkemeier;alkaliattachedtoheliumclusters}
\bibinfo{author}{\bibfnamefont{F.}~\bibnamefont{Stienkemeier}},
  \bibinfo{author}{\bibfnamefont{J.}~\bibnamefont{Higgins}},
  \bibinfo{author}{\bibfnamefont{C.}~\bibnamefont{Callegari}},
  \bibinfo{author}{\bibfnamefont{S.~I.} \bibnamefont{Kanorsky}},
  \bibinfo{author}{\bibfnamefont{W.~E.} \bibnamefont{Ernst}}, \bibnamefont{and}
  \bibinfo{author}{\bibfnamefont{G.}~\bibnamefont{Scoles}},
  \bibinfo{journal}{Z. Phys. D} \textbf{\bibinfo{volume}{38}},
  \bibinfo{pages}{253} (\bibinfo{year}{1996}).

\bibitem[{\citenamefont{Reho et~al.}(2000)\citenamefont{Reho, Higgins,
  Callegari, Lehmann, and Scoles}}]{Reho;AlkaliExciplexForationI}
\bibinfo{author}{\bibfnamefont{J.}~\bibnamefont{Reho}},
  \bibinfo{author}{\bibfnamefont{J.}~\bibnamefont{Higgins}},
  \bibinfo{author}{\bibfnamefont{C.}~\bibnamefont{Callegari}},
  \bibinfo{author}{\bibfnamefont{K.~K.} \bibnamefont{Lehmann}},
  \bibnamefont{and} \bibinfo{author}{\bibfnamefont{G.}~\bibnamefont{Scoles}},
  \bibinfo{journal}{J. Chem. Phys.} \textbf{\bibinfo{volume}{113}},
  \bibinfo{pages}{9686} (\bibinfo{year}{2000}).

\bibitem[{\citenamefont{Bruhl et~al.}(2001)\citenamefont{Bruhl, Trasca, and
  Ernst}}]{Bruhl;RbHeexiFormationDroplets}
\bibinfo{author}{\bibfnamefont{F.~R.} \bibnamefont{Bruhl}},
  \bibinfo{author}{\bibfnamefont{R.~A.} \bibnamefont{Trasca}},
  \bibnamefont{and} \bibinfo{author}{\bibfnamefont{W.~E.} \bibnamefont{Ernst}},
  \bibinfo{journal}{J. Chem. Phys.} \textbf{\bibinfo{volume}{115}},
  \bibinfo{pages}{10220} (\bibinfo{year}{2001}).

\bibitem[{\citenamefont{Schulz et~al.}(2001)\citenamefont{Schulz, Claas, and
  Stienkemeier}}]{Schulz;FormationExciplexesKHe}
\bibinfo{author}{\bibfnamefont{C.}~\bibnamefont{Schulz}},
  \bibinfo{author}{\bibfnamefont{P.}~\bibnamefont{Claas}}, \bibnamefont{and}
  \bibinfo{author}{\bibfnamefont{F.}~\bibnamefont{Stienkemeier}},
  \bibinfo{journal}{Phys. Rev. Lett.} \textbf{\bibinfo{volume}{87}},
  \bibinfo{pages}{153401} (\bibinfo{year}{2001}).

\bibitem[{\citenamefont{Kanorsky et~al.}(1994)\citenamefont{Kanorsky, Arndt,
  Dziewior, Weis, and H{\"a}nsch}}]{Kanorsky;OptSpecAtomsTrappedSolidHe}
\bibinfo{author}{\bibfnamefont{S.~I.} \bibnamefont{Kanorsky}},
  \bibinfo{author}{\bibfnamefont{M.}~\bibnamefont{Arndt}},
  \bibinfo{author}{\bibfnamefont{R.}~\bibnamefont{Dziewior}},
  \bibinfo{author}{\bibfnamefont{A.}~\bibnamefont{Weis}}, \bibnamefont{and}
  \bibinfo{author}{\bibfnamefont{T.~W.} \bibnamefont{H{\"a}nsch}},
  \bibinfo{journal}{Phys. Rev. B} \textbf{\bibinfo{volume}{49}},
  \bibinfo{pages}{3645} (\bibinfo{year}{1994}).

\bibitem[{\citenamefont{Nettels et~al.}(2003)\citenamefont{Nettels,
  M\"{u}ller-Siebert, Ulzega, and Weis}}]{Nettels;MultiPhoton1}
\bibinfo{author}{\bibfnamefont{D.}~\bibnamefont{Nettels}},
  \bibinfo{author}{\bibfnamefont{R.}~\bibnamefont{M\"{u}ller-Siebert}},
  \bibinfo{author}{\bibfnamefont{S.}~\bibnamefont{Ulzega}}, \bibnamefont{and}
  \bibinfo{author}{\bibfnamefont{A.}~\bibnamefont{Weis}},
  \bibinfo{journal}{Applied Physics B} \textbf{\bibinfo{volume}{77}},
  \bibinfo{pages}{563} (\bibinfo{year}{2003}).

\bibitem[{\citenamefont{Kinoshita et~al.}(1995)\citenamefont{Kinoshita, Fukuda,
  Takahashi, and Yabuzaki}}]{OpitcalPropAlkaliAtomsPressSuperfluidHe;Kinoshita}
\bibinfo{author}{\bibfnamefont{T.}~\bibnamefont{Kinoshita}},
  \bibinfo{author}{\bibfnamefont{K.}~\bibnamefont{Fukuda}},
  \bibinfo{author}{\bibfnamefont{Y.}~\bibnamefont{Takahashi}},
  \bibnamefont{and} \bibinfo{author}{\bibfnamefont{T.}~\bibnamefont{Yabuzaki}},
  \bibinfo{journal}{Phys. Rev. A} \textbf{\bibinfo{volume}{52}},
  \bibinfo{pages}{2707} (\bibinfo{year}{1995}).

\bibitem[{\citenamefont{Nettels et~al.}()\citenamefont{Nettels, Hofer,
  Moroshkin, M\"{u}ller-Siebert, Ulzega, and Weis}}]{Nettels;ExciplexesPhd}
\bibinfo{author}{\bibfnamefont{D.}~\bibnamefont{Nettels}},
  \bibinfo{author}{\bibfnamefont{A.}~\bibnamefont{Hofer}},
  \bibinfo{author}{\bibfnamefont{P.}~\bibnamefont{Moroshkin}},
  \bibinfo{author}{\bibfnamefont{R.}~\bibnamefont{M\"{u}ller-Siebert}},
  \bibinfo{author}{\bibfnamefont{S.}~\bibnamefont{Ulzega}}, \bibnamefont{and}
  \bibinfo{author}{\bibfnamefont{A.}~\bibnamefont{Weis}}, \bibinfo{note}{to be
  submitted}.

\bibitem[{\citenamefont{Pascale}(1983)}]{Pascale;LDependentPseudoPotentialAlka%
liHe}
\bibinfo{author}{\bibfnamefont{J.}~\bibnamefont{Pascale}},
  \bibinfo{journal}{Phys. Rev. A} \textbf{\bibinfo{volume}{28}},
  \bibinfo{pages}{632} (\bibinfo{year}{1983}).

\bibitem[{\citenamefont{Beck}(1968)}]{Beck;InteratomicPotHeHe}
\bibinfo{author}{\bibfnamefont{D.~E.} \bibnamefont{Beck}},
  \bibinfo{journal}{Molecular Physics} \textbf{\bibinfo{volume}{14}},
  \bibinfo{pages}{311} (\bibinfo{year}{1968}).

\bibitem[{\citenamefont{Nettels}(2003)}]{Nettels;Phd}
\bibinfo{author}{\bibfnamefont{D.}~\bibnamefont{Nettels}},
  \bibinfo{type}{Ph.{D}. thesis}, \bibinfo{school}{University Fribourg,
  Switzerland} (\bibinfo{year}{2003}).

\bibitem[{\citenamefont{Stienkemeier}()}]{Stienkemeier;CsPi12He2}
\bibinfo{author}{\bibfnamefont{F.}~\bibnamefont{Stienkemeier}},
  \bibinfo{note}{private communication}.

\end{thebibliography}

\end{document}